\documentclass[aps,pre,twocolumn,a4paper,superscriptaddress,floatfix,10pt,longbibliography]{revtex4-2}

\usepackage[T1]{fontenc}
\usepackage{graphicx}
\usepackage{dcolumn}
\usepackage{bm}
\usepackage{color}
\usepackage{amsmath}
\usepackage{amssymb}
\usepackage{hyperref}
\usepackage{subfigure}
\usepackage{float}
\usepackage{bigints}
\usepackage{changes}
\newcommand{\dd}{{\mathrm{d}}}
\newcommand{\gs}{{\mathrm{gs}}}
\newcommand{\mm}{{\mathrm{min}}}
\newcommand{\accnn}{{\overline{\text{CNN}}}}
\begin{document}

\title{Deep learning density functionals for gradient descent optimization}

\author{E. Costa}
\affiliation{School of Science and Technology, Physics Division, Universit\`a di Camerino, 62032 Camerino, Italy}
\affiliation{INFN-Sezione di Perugia, 06123 Perugia, Italy}

\author{G. Scriva}
\affiliation{School of Science and Technology, Physics Division, Universit\`a di Camerino, 62032 Camerino, Italy}
\affiliation{INFN-Sezione di Perugia, 06123 Perugia, Italy}

\author{R. Fazio}
\affiliation{Abdus Salam ICTP, Strada Costiera 11, I-34151 Trieste, Italy}
\affiliation{Dipartimento di Fisica, Universit\`a di Napoli ``Federico II'', Monte S. Angelo, I-80126 Napoli, Italy}

\author{S. Pilati}
\affiliation{School of Science and Technology, Physics Division, Universit\`a di Camerino, 62032 Camerino, Italy}
\affiliation{INFN-Sezione di Perugia, 06123 Perugia, Italy}

\begin{abstract}
Machine-learned regression models represent a promising tool to implement accurate and computationally affordable energy-density functionals to solve quantum many-body problems via density functional theory. However, while they can easily be trained to accurately map ground-state density profiles to the corresponding energies, their functional derivatives often turn out to be too noisy, leading to instabilities in self-consistent iterations and in gradient-based searches of the ground-state density profile.
We investigate how these instabilities occur when standard deep neural networks are adopted as regression models, and we show how to avoid them by using an ad-hoc convolutional architecture featuring an inter-channel averaging layer. The main testbed we consider is  a realistic model for noninteracting atoms in optical speckle disorder.
With the inter-channel average, accurate and systematically improvable ground-state energies and density profiles are obtained via gradient-descent optimization, without instabilities nor violations of the variational principle.

\end{abstract}

\maketitle

\section{Introduction}
Density functional theory (DFT) is the workhorse of computational material science and  quantum chemistry~\cite{doi:10.1063/1.4704546}. It is based on  rigorous theorems~\cite{PhysRev.136.B864,doi:10.1073/pnas.76.12.6062,Lieb2002} certifying that the ground-state energy can be computed minimizing a (generally unknown) functional of the density profile, allowing bypassing computationally prohibitive wave-function based methods.
However, the available approximations for the  density functional are reliable only for weakly correlated materials, while in the regime of strong electron correlations dramatic failures may occur~\cite{cohen2012challenges}.
In recent years, machine-learning (ML) algorithms have reached remarkable breakthroughs in various branches of physics research~\cite{Dunjko_2018,carleo2019machine,doi:10.1080/23746149.2020.1797528,10.1088/2516-1075/ac572f},  and they have also been adopted in the framework of DFT, both for continuous-space~\cite{PhysRevLett.108.253002,https://doi.org/10.1002/qua.25040,Brockherde2017,PhysRevA.100.022512,doi:10.1126/science.abj6511} and for one-dimensional tight-binding models~\cite{PhysRevB.99.075132,PhysRevLett.125.076402,PhysRevResearch.2.033388}.
These algorithms pave the way to data based  approaches to the development of density functionals. Furthermore, they facilitate the implementation of computationally convenient strategies based on orbital-free DFT~\cite{Meyer2020,2022}.
Most previous studies adopted relatively simple regression models, such as, e.g., kernel ridge regression, showing that moderately large training sets of ground-state density profiles and corresponding energies allow reconstructing remarkably accurate density functionals. 
Also artificial neural networks have been adopted (see, e.g., \cite{PhysRevA.100.022512,2022}), considering standard architectures such as the convolutional neural networks (CNNs) popular in the field of computer vision.
Unfortunately, in the case of continuous-space models, severe drawbacks have emerged when such ML functionals have been employed in self-consistent calculations and in gradient-based optimizations. Specifically, the functional derivatives turned out to be too noisy, leading to unphysical density profiles and to strong violation of the variational principle~\cite{PhysRevLett.108.253002,https://doi.org/10.1002/qua.24937,2022}.
Some remedial strategies have already been explored. Essentially, they resort to gradient denoising via dimensionality reduction~\cite{https://doi.org/10.1002/qua.24937} or basis truncation~\cite{PhysRevB.94.245129}, to constrained optimization, or they aim at exploiting additional information (e.g., energy derivatives) in the training process~\cite{doi:10.1126/science.abj6511,2022}. These strategies have provided significant benefits, but they have some limitations, as they might lead to variational biases or they require additional data which is far less accessible.
Due to the pivotal role played by DFT, further complementary strategies are highly desirable.

In this Article, we investigate the use of deep neural networks as regression models to reconstruct continuous-space density functionals from training data. 
Our main finding is that a tailored convolutional network featuring an inter-channel averaging operation  allows avoiding the drawbacks mentioned above.
Following analogous previous studies ~\cite{PhysRevLett.108.253002,https://doi.org/10.1002/qua.25040,Meyer2020,2022,https://doi.org/10.1002/qua.24937},  the testbed we consider is a single particle model, but we mostly focus on a more realistic Hamiltonian which describes ultracold atoms moving in one-dimensional optical speckle patterns. 
Our analysis is complemented by addressing deep-well models from the literature (see Appendix).
Our aim is to develop a sufficiently effective deep-learned functional to allow  searching the ground-state energy and the density profile of previously unseen instances of speckle disorder via gradient-descent optimization.
We show that the most popular network architectures, namely, the standard CNNs, are inadequate for this task. Indeed, while they provide remarkably accurate energy predictions when fed with exact ground-state density profiles, their functional derivatives are too noisy. This leads to instabilities in the gradient-descent search of the density profile which minimizes the energy functional, unless the accuracy is jeopardized via an early halting of the optimization procedure. 
We demonstrate that these instabilities can be avoided with the tailored neural network. This is inspired by an ensemble-averaging mechanism, and it features, beyond the standard multi-channel convolutional layers, additional layers that perform an inter-channel averaging operation. 
We show that this feature allows us iterating gradient-descent steps at will, providing accurate results that can be systematically improved by increasing the number of channels.

The rest of the article is organized as follows:
in Section~\ref{sectionmethods} we describe the formalism of DFT based on deep learning, the structure of the standard and of the average-channel neural networks, as well as the gradient-descent technique used to find ground-state energies and densities.
The main testbed model we address is described in Section~\ref{modeldataset}. Therein we also report details on the dataset, on the protocol used for network training, and on the  accuracy reached in the regression task.
The results obtained in gradient-descent optimization with the standard and with the average-channel neural networks are compared in Section~\ref{Results}. The instabilities occurring with standard networks are highlighted, and their suppression with the inclusion of the average layer is discussed in detail.
Section~\ref{conclusions} provides a summary of the main findings and some comments on future perspectives.
To favor comparison with previous studies~\cite{PhysRevLett.108.253002,Meyer2020}, in the Appendix~\ref{appendix} we report the test of the average-channel neural network in deep potential wells defined by the sum of three Gaussian functions.

\section{Density function theory with artificial neural networks}
\label{sectionmethods}

ML provides novel promising approaches to learn energy-density functionals for DFT from data. These functionals have the potential to accurately describe strongly correlated systems. However, their variational minimization to search for the ground-state density profile turned out to be problematic due to noisy functional derivatives ~\cite{https://doi.org/10.1002/qua.24937}, ~\cite{https://doi.org/10.1002/qua.25040}, ~\cite{PhysRevLett.108.253002}. This 
problem is already evident in noninteracting systems. Henceforth, in the following we focus on single-particle problems, but the technique we develop can be applied to interacting systems via the creation of suitable training sets.
%

In this article we consider one-dimensional  single-particle Hamiltonians written in the form:
\begin{equation}
\label{H}
    H=-\frac{\hbar^2}{2m}\frac{\dd^2}{\dd x^2} + V(x),
\end{equation}
where $\hbar$ is the reduced Planck constant and $m$ is the particle mass. The external potential $V(x)$ is compatible with the adopted periodic boundary conditions.
In the framework of DFT, one aims at computing the ground-state energy $e_\gs$ of the Hamiltonian~\eqref{H} from a functional of the density profile: $e_\gs = E[n_\gs]$.
Here, $n_\gs$ indicates the density profile
\begin{equation}
\label{density}
    n_\gs(x)=\left|\psi_\gs(x)\right|^2,
\end{equation}
where $\psi_\gs(x)$ is the ground-state wave function.
The first Hohenberg-Kohn theorem guarantees that, in principle, this functional exists~\cite{PhysRev.136.B864}.
In practice, it is convenient to separate the known potential energy contribution, seeking for a functional for the kinetic energy only\footnote{In interacting systems, the unknown functional would include also correlation effects, while the mean-field contribution would be conveniently encoded in the so-called Hartree term.}:
\begin{equation}
\label{kin}
    t_\gs =  e_\gs - \int_0^L \dd x  n_\gs(x) V(x) \equiv T[n_\gs].
\end{equation}
It is worth pointing out that we do not adopt the Kohn-Sham formalism. As in previous studies on ML-based DFT ~\cite{Meyer2020,2022}, ~\cite{https://doi.org/10.1002/qua.24937}, ~\cite{https://doi.org/10.1002/qua.25040}, ~\cite{PhysRevLett.108.253002}, the orbital-free formalism is used, attempting to approximate the kinetic energy (eventually together with energy terms in interacting systems) as a functional of the density. If successful, this attempt would therefore also lead to a significant reduction of computational cost compared to the more demanding Kohn-Sham approach.

Deep-learning techniques can be adopted in the DFT framework. The first  task is to train a deep neural network to map ground-state density profiles to the corresponding kinetic energies, therefore learning the unknown functional $T[n_\gs]$.
This can be achieved via supervised learning from a dataset including many instances of density profiles associated to the corresponding kinetic energies, $\left\{ n_{\gs,k}, t_{\gs,k} \right\}$. The integer $k$ labels the instances in the dataset.
The parameters of the neural network, collectively denoted with $\omega$, are optimized by minimizing the loss function, namely, the mean squared error
\begin{equation}
\label{MSE}
    \mathcal{L}(\omega)=\frac{1}{N_{\mathrm{train}}} \sum_{k=1}^{N_{\mathrm{train}}} \left| t_{\gs,k}-\tilde{T}_{\gs,k}(\omega)[n_{\gs,k}]\right|^2,
\end{equation}
where $\tilde{T}_{\gs,k}(\omega)[n_{\gs,k}]$ denotes the kinetic energy predicted by the neural network and $N_{\mathrm{train}}$ is the number of instances in the training set.  This optimization can be performed using the stochastic gradient descent algorithms or one of its many successful variants.
%

\subsection{Neural networks}
The first regression model we consider is a standard CNN. Its structure is familiar from many fields where deep learning has proven successful such as, e.g,  image recognition.
It is composed of $N_b=3$ convolutional blocks. Each block includes a convolutional layer with $N_c$ channels (this  hyperparameter is specified in Table~\ref{table1}), whose  filter size is $k_f=13$ and padding type is periodic, and an average pooling layer with a kernel size $k_s=2$. Two variants of this network are considered, using two popular activation functions, namely the ReLU  function, defined as
\begin{equation}
    \mathrm{ReLU}(x)=\begin{cases}
    x & \text{if } x>0 \\
    0 & \text{otherwise} \\
    \end{cases},
\end{equation}
and the Softplus function 
\begin{equation}
    \mathrm{Softplus}(x)=\ln{(1+\exp(x))}.
\end{equation}
The last convolutional block is processed through a flattening layer and then connected to a dense layer with only one neuron (identity activation function) to generate a scalar output.
As discussed in detail in  Section~\ref{Results}, these standard CNNs turn out to be inadequate for the DFT framework.
In particular, their functional derivatives are too noisy to perform a gradient-based search of the ground state.
Therefore, we introduce a novel tailored architecture which features an inter-channel average operation.
In the following, this network will be referred to as average-channel CNN ($\accnn$). 
Specifically, this model is composed of two convolutional blocks, each including $N_c$ convolutional channels ($k_f=13$ and periodic padding),
an average pooling layer ($k_s=4$ in the first block and $k_s=2$ in the second block), and, notably, an additional layer, where each neuron computes the average of the activations of the corresponding neurons in all channels of the previous layer. 
The activation function is ReLU. The last convolutional block passes a flatten layer and is then connected to one dense layer with only one neuron, as in the standard CNN case.
It is worth pointing out that this average operation reduces the scaling of the number of parameters from quadratic to linear in $N_c$. This allows considering architectures with many channels without facing prohibitive computational costs nor overfitting problems.
In each architecture adopted in this article, all convolutional blocks feature the same number of channels. Explorations performed with different numbers lead to similar findings, so we do not discuss them to avoid burdening the presentation. 
Hereafter, we describe the operations performed by the convolutional blocks more formally.

In a standard CNN, the action of the $n$-th convolutional block corresponds to the following convolution operation:
\begin{equation}
\label{refconv1}
\begin{split}
    h^{\alpha}_{n}(x)=\frac{1}{k_s}\int^{x+k_s/2}_{x-k_s/2} \dd y \; \\
    \mathrm{act} \left[ \sum_{\beta} \int_{y-k_f/2}^{y+k_f/2} \dd x' \; W^{\alpha \beta}_{n}(y,x') h^{\beta}_{n-1}(x') +v^{\alpha}_{n} \right] &,
\end{split}
\end{equation}
where $\mathrm{act}$ is the chosen activation function, the matrices $W^{\alpha \beta}_{n}$ represent, for each filter $\alpha$ and input channel $\beta$, a kernel of size $ k_f $, and $v^{\alpha}$ are the set of biases for each filter.
Instead, in the $\accnn$, the $n$-th block has an additional inter-channel average operation, which is expressed as:
\begin{equation}
\label{refconv2}
\begin{split}
    \Bar{h}_{n}(x)=\frac{1}{N_{c}} \sum_{\alpha} h^{\alpha}_{n}(x), 
\end{split}
\end{equation}
where
\begin{equation}
\label{eqref2}
\begin{split}
    h^{\alpha}_{n}(x)= \frac{1}{ k_s}\int^{x+ k_s/2}_{x- k_s/2} dy \; \\
    \mathrm{act} \left[ \int_{y-k_f/2}^{y+k_f/2} \dd x' \; W^{\alpha}_{n}(y,x') \Bar{h}_{n-1}(x') + v^{\alpha}_{n} \right] 
\end{split}
\end{equation}
represents the previous $N_{c}$ parallel convolution operations. 
Notice that in Eqs.~\eqref{refconv1} and ~\eqref{refconv2} integrals actually indicate discrete operations.
All the neural networks considered in this work are implemented and trained using the Pytorch library, exploiting automatic differentiation to compute discrete functional derivatives~\cite{paszke2017automatic}.



\subsection{Formalism of gradient-descent optimization}
Once the kinetic energy functional  has been learned by the neural network, and consequently we assume $T[n] \equiv \tilde{T}{_\omega}[n]$, both the ground-state energy and the density  corresponding to a new instance of the Hamiltonian can be obtained from the variational principle. 
Indeed, the second Hohenberg-Kohn theorem ensures that the (exact) functional is minimized by the ground-state density profile~\cite{PhysRev.136.B864}.
This can be expressed using the Euler equation:
\begin{equation}
    \frac{\delta T[n]}{\delta n(x)} + V(x) -\mu =0
\end{equation}
where $\mu$ is a normalization constraint. 
Its solution can be efficiently obtained using the gradient-descent algorithm, as usually done in orbital-free DFT~\cite{orbitalfree}.
Specifically, one iterates the following update rule:
\begin{equation}
    n_{t+1}(x)= n_{t}(x) - \eta \left( \frac{\delta T[n_t]}{\delta n(x)} +V(x) - \mu_{t} \right),
    \label{n gradient descent}
\end{equation}
starting from a reasonably chosen initial profile $n_0(x)$. In this equation, $\eta>0$ is the chosen learning rate, the integer $t=0,1,2,\dots,t_{\mathrm{max}}$ labels the steps, and the adaptive coefficient $\mu_{t}$ is introduced to ensure the normalization condition: %
\begin{equation}
    \int \dd x n_t(x) = 1.
\end{equation}
To ensure the density never becomes negative, it is convenient to perform a variable change in Eq.~\eqref{n gradient descent}: $\chi= \sqrt{n(x)}$. This leads to the update rule:
\begin{equation}
\label{step}
   \chi_{t+1}(x)=\chi_{t}(x) - \eta \left( \frac{\delta T[\chi^2_t]}{\delta \chi(x)} + 2\chi_t(x) V(x) - 2 \chi_t(x) \mu_t \right),
\end{equation}
where the normalization coefficient is computed as:
\begin{equation}
\label{mut}
    \mu_t=\frac{\displaystyle\int \dd x \left( \frac{1}{2} \frac{\delta T[\chi^2_t]}{\delta \chi_t(x)} \chi_t(x) +\chi^2_t(x) V(x) \right)}{\displaystyle\int \dd x \; \chi^2_t (x)}.
\end{equation}
%
%
%
The computation of the square of the function $\chi$ is performed by an additional layer in Pytorch. Henceforth, the functional derivative with respect to $\chi$ can be directly computed exploiting automatic differentiation.
%


Whether the gradient descent algorithm reaches the ground-state or not depends on two major issues. First, the optimization might get stuck in a local minimum. Indeed, the optimization landscape is not proven to be convex, even for the (unknown) exact functional. Convexity can be instead proven for an extended functional, defined in a domain including density profiles not corresponding to ground states~\cite{Lieb2002}.
This problem can be 
mitigated by repeated the minimization process starting from different initial profiles, or by introducing random steps based on, e.g., Metropolis-type algorithms.
The second issue is the accuracy of the functional derivative. Noisy and  inaccurate derivatives might create unphysical density profiles, 
clearly not corresponding to ground states. Since the regression model was not trained on such profiles, it might provide very inaccurate energy predictions, even lower than the exact ground-state energy $e_\gs$. This leads to dramatic failures of the gradient descent optimization, even to large violations of the variational principle.
This problem has already been emphasized in the literature, and it was indicated as a major challenge to be overcome for the further development of ML based DFTs.
In Refs.~\cite{https://doi.org/10.1002/qua.24937},  \cite{PhysRevLett.108.253002} \cite{https://doi.org/10.1002/qua.25040}, the adopted regression model was kernel ridge regression. 
This model typically requires smaller datasets for training, but it is less efficient  than the deep neural networks in systematically extracting further information from larger and larger datasets. Even kernel ridge regression led to noisy derivatives. 
To circumvent this problem, the authors introduced two main techniques, referred to as local principal component analysis and non linear gradient denoising. They aim at projecting the functional derivative to the manifold tangent to the one spanned by ground-state density profiles.
Other works included derivative data in the training process -- also adopting standard CNNs --- using the Sobolev Loss~\cite{Meyer2020,2022}.
We emphasize that our goal is to train the regression model using only ground-state density profiles and the corresponding energies, avoiding resorting to less accessible data such as exited-state properties or energy gradients.


\begin{table}[h!]
\begin{tabular}{c c c c c c} 
 \hline
 \hline 
 \rule{0pt}{3ex}\noindent
 Blocks & Activation & $N_c$ & $k_s$ & Neural network & $R^2$ \\ [1ex]
 \hline
 \rule{0pt}{3ex}\noindent
 $2$ & ReLU     & $60$   & $[4,2]$ &  $\accnn$ & $0.99961$ \\ 
 $2$ & ReLU     & $140$  & $[4,2]$ &  $\accnn$ & $0.99969$ \\
 $2$ & ReLU     & $260$  & $[4,2]$ &  $\accnn$ & $0.99967$ \\
 $3$ & ReLU     & $30$   & $2$ &  CNN   & $0.99996$ \\
 $3$ & Softplus & $30$   & $2$ &  CNN   & $0.99991$ \\
 \hline
 \hline
\end{tabular}
\caption{Coefficient of determination $R^2$ for two standard CNNs with ReLU and with Softplus activation functions, and for three $\accnn s$ with different number of channels $N_c$. 
The test set includes $15000$ (previously unseen) instances of the optical speckle pattern. The first column reports the number of convolutional blocks.}
\label{table1}
\end{table}

\section{Testbed model and training dataset}
\label{modeldataset}
The main testbed model we consider is the single-particle model~\eqref{H}, where the (random) external potential $V(x)$ is designed to represent the effect of optical speckle patterns on ultracold atoms. 
Notice that another testbed, borrowed from the literature, is considered in the Appendix~\ref{appendix}, allowing us to further characterize the domain of applicability of the $\accnn$.
The speckle potentials can be created by applying a specific filter in Fourier space to a random complex gaussian field ~\cite{goodman2007speckle}. The filter corresponds to the aperture of the optical apparatus used to experimentally create the field, and it fixes the characteristic size of the speckle grains. 
In fact, with this choice the Hamiltonian~\eqref{H} describes early cold-atom experiments on Anderson localization in one dimension~\cite{billy2008direct,roati2008anderson}.
The statistical and the spectral properties of optical speckle patterns are known~\cite{goodman2007speckle,PhysRevA.94.022114,pilati2019supervised}.
The intensity of the potential $V$ in a point $x$ follows the probability distribution
\begin{equation}
    P(V)=\exp \left( -\frac{V}{V_0} \right),
\end{equation}
for $V \ge 0$, and $P(V)=0$ otherwise; $V_0\ge 0$ is the average intensity, and it also coincides with the standard deviation.
It is the unique parameter determining the disorder strength.
The two-point autocorrelation function satisfies the following equation:
\begin{equation}
    \frac{\langle V(x' +x)V(x')\rangle}{V^{2}_0} -1=\frac{ \sin(\pi x/ \gamma)^2}{(\pi x / \gamma)^2},
\end{equation}
where $\gamma$ determines the correlation length, namely, the size of the typical speckle grains.
In the above equation, the brackets $\langle \cdot \rangle$  indicate the average over many random realizations of the speckle pattern. This ensemble average coincides with the spatial average for sufficiently large systems.
The correlation energy $E_c=\frac{\hbar^2}{2m \gamma}$ separates the strong disorder regime $V_0 \gg E_c$, where the low-energy orbitals are localized on a length scale of order $\gamma$~\cite{PhysRevA.100.013603} due to the Anderson localization phenomenon~\cite{PhysRevLett.42.673}, from the weak disorder regime $V_0 \ll E_c$, where their localization length is much larger. 
Notice that in one dimension any disorder strength induces Anderson localization in a sufficiently large system~\cite{anderson1958absence}.
In the following, we consider the relatively large system size $L=14 \gamma$ and the intermediate disorder strength $V_0=0.5 E_c$. 
This choice allows generating rather variegate ground-state density profiles with different shapes and varying degrees of localization, depending on the details of the specific realization of the speckle pattern. Therefore, the Hamiltonian~\eqref{H} represents a stringent testbed for the DFT framework.
Two representative instances of the speckle pattern are shown in Fig.~\ref{fig1}, together with the corresponding ground-state density profiles.
%

Different random realizations of the speckle pattern can be efficiently generated on a discrete grid with the algorithm described in Refs.~\cite{huntley1989speckle,PhysRevA.73.013606}. We choose a fine grid with $N_g=256$ points, such that the grid step $\delta_x = L/N_g \ll \gamma$.
%
The ground-state energy $e_\gs$ and the corresponding orbital $\psi_\gs(x)$ are determined via exact diagonalization using a high-order finite difference formula. Computations performed with finer grids show that the discretization error is negligible.
The choice of such a fine grid allows us computing all spatial integrals (see, e.g, those in Eqs. \eqref{kin} and \eqref{mut}) with the discrete approximation $\int_0^L \dd x \longrightarrow \delta_x \sum^{N_{g}}_{i=1}$. Higher order approximations lead to essentially indistinguishable results for our purposes.
Furthermore, the functional derivative in Eq.~\eqref{step} is computed as
\begin{equation}
    \frac{\delta T[\psi]}{\delta \psi(x)} = \frac{\partial T[\psi]}{\partial \psi(x_i)} \frac{1}{\delta_x}.
\end{equation}
For this, we exploit Pytorch automatic differentiation.
The training of the neural networks, namely, the minimization of the loss function Eq.~\eqref{MSE}, is performed using the Adam algorithm~\cite{kingma2014adam}. 
The chosen learning rate is $l_r=10^{-4}$, the minibatch size is $N_b=100$ and the training epochs are $N_e=1200$.
The other parameters of the Adam algorithm are set at their suggested default values.
%
Our global  dataset is composed of $150000$ instances. As customary in deep-learning studies, we split it into a training set ($81 \%$), a validation set ($9 \%$), and a test set ($10 \%$).
%
To measure the accuracy in the regression task, we consider the coefficient of determination, defined as:
\begin{equation}
\label{R2}
    R^2= 1- 
    \frac{
    \sum_{k=1}^{N_{\mathrm{test}}}
    \left|t_{\gs,k}-\tilde{T}_{\omega}[n_{\gs,k}]\right|^2}
    {N_{\mathrm{test}}\sigma^2},
\end{equation}
where $N_{\mathrm{test}}$ is the number of instances in the test set and $\sigma^2=\frac{1}{N_{\mathrm{test}}}\sum_{k=1}^{N_{\mathrm{test}}} \left(t_{\gs,k} - \bar{t}_{\gs}\right)^2$ is the variance of their kinetic energies; with $\bar{t}_{\gs}=\frac{1}{N_{\mathrm{test}}}\sum_{k=1}^{N_{\mathrm{test}}} t_{\gs,k}$ we denote the average kinetic energy.
After training, the two variants of standard CNNs reach remarkable accuracies on the test set, meaning that, when they are provided with an exact ground-state density profile corresponding to a previously unseen speckle pattern, they accurately predict the associated ground-state kinetic energy and, via Eq.~\eqref{kin}, also the total energy. The $R^2$ scores obtained with these two CNNs are reported in Table~\ref{table1}. 
%
%
The $R^2$ scores reached by the $\accnn$ are comparable, but slightly inferior, to the ones obtained by the standard CNNs (see Table~\ref{table1}).
Remarkably, despite of this (slightly) lower performance in kinetic energy predictions, the averaging operation drastically suppresses the noise in the functional derivative, allowing the use of $\accnn$ in a gradient-based search of the ground-state energy and density profile. This is discussed in Section~\ref{Results}.

\section{Results for gradient-descent optimization}
\label{Results}

\begin{figure}[h!]

  \centering
  \subfigure{\includegraphics[scale=0.15]{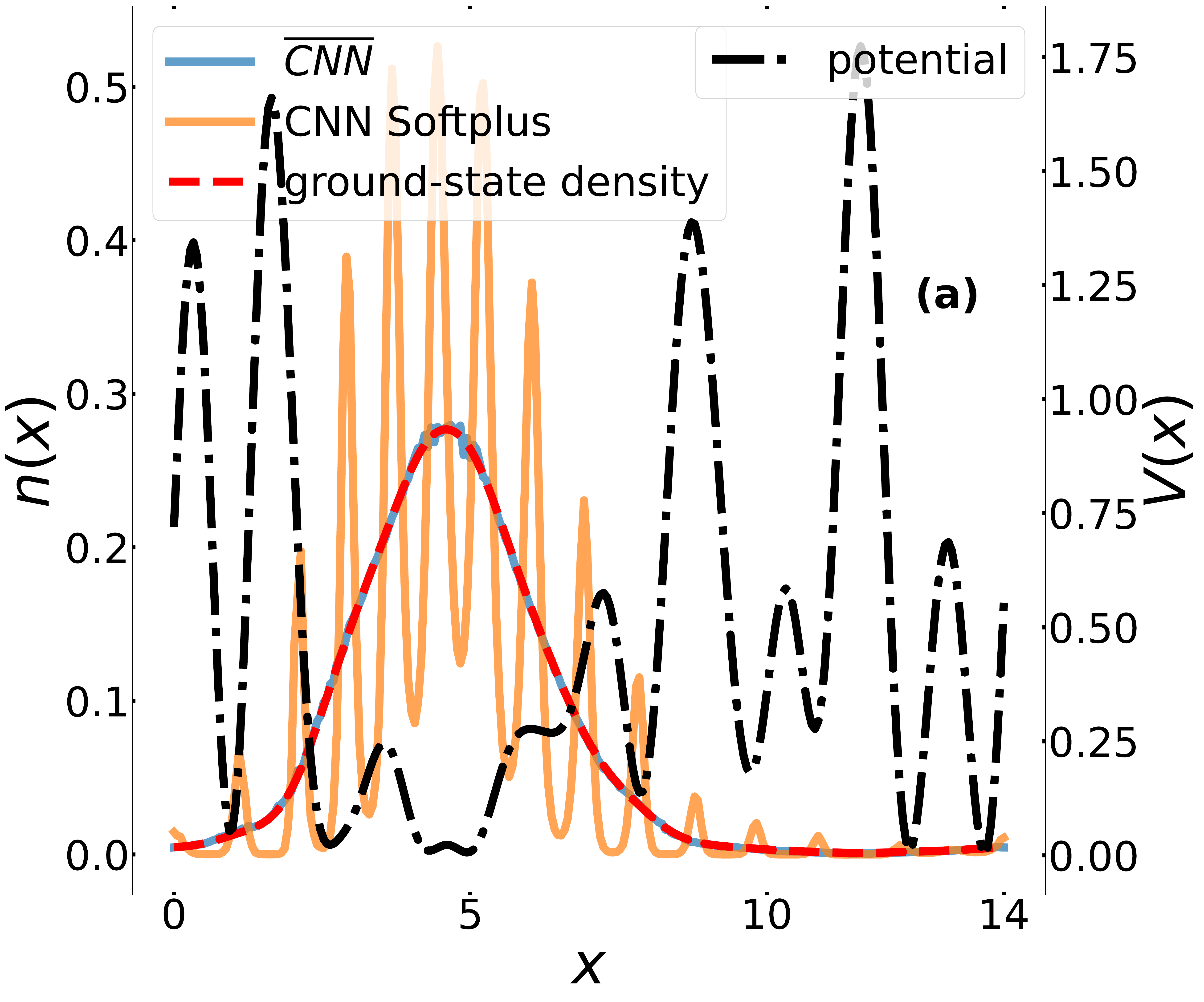} }
  \subfigure{\includegraphics[scale=0.15]{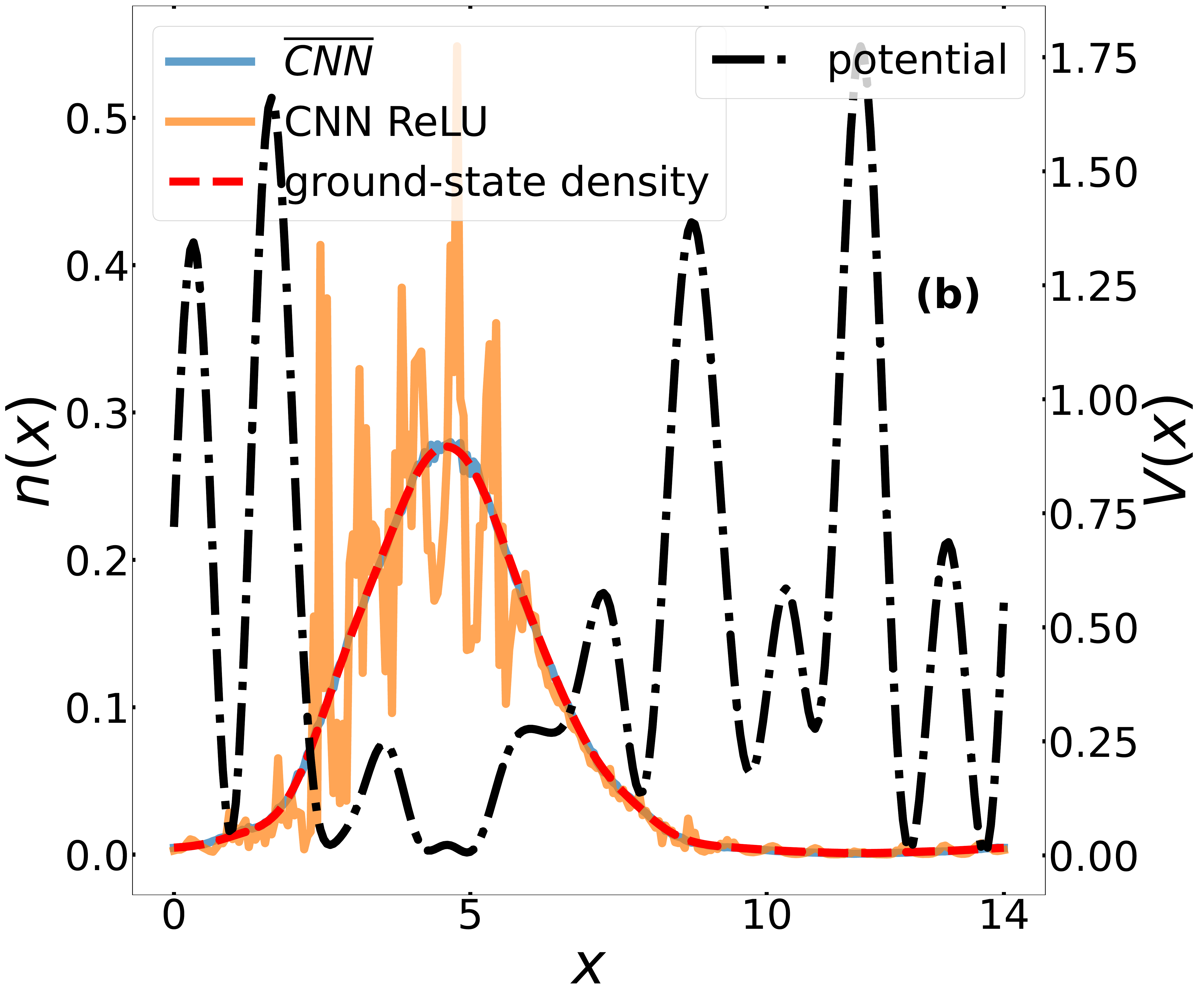}}
 \caption{
 Panels (a) and (b) show the external potentials $V(x)$ representing two instances of the optical speckle pattern (right vertical axis, the unit is the correlation energy $E_c$) and the corresponding ground-state density profiles (left vertical axis, units of $1/\gamma$). The spatial variable $x$ is in units of the correlation length $\gamma$.
 In panel (a) the exact ground state is compared to the DFT result obtained using the standard CNN with Softplus activation function and using the $\accnn$.
 In panel (b) the standard CNN with ReLU activation is considered instead.
 }
 \label{fig1}
\end{figure}

\begin{figure}[h!]
  \centering
  \subfigure{\includegraphics[scale=0.3]{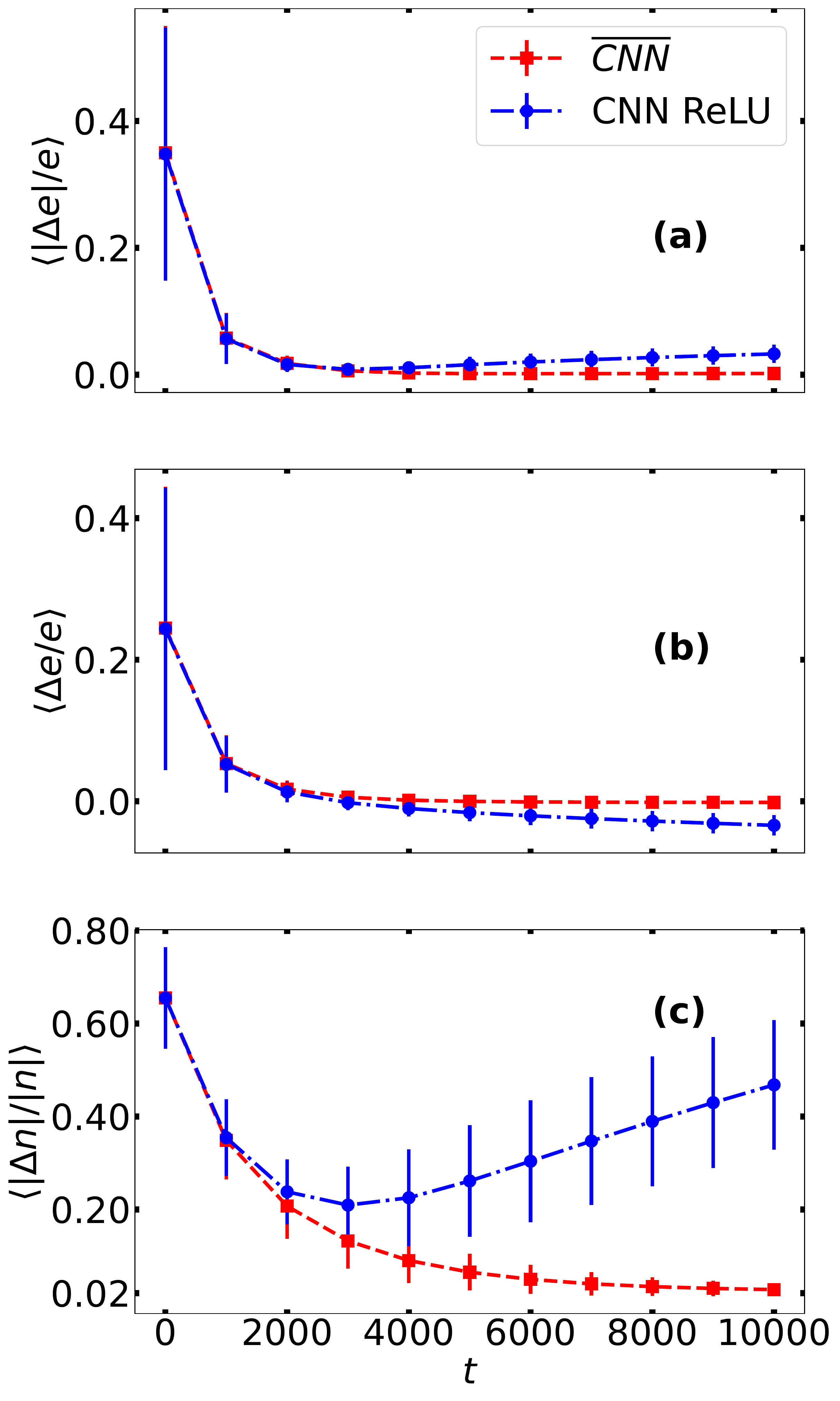}}
  \caption{
Relative discrepancies of DFT predictions from exact grounds-state results as a function of the number of steps $t$ of the gradient-descent optimization.
The results are averaged over a test set of 500 speckle-pattern instances. Three error metrics are shown: absolute energy discrepancy $\left< |\Delta e| /e \right>$ [panel (a)], energy discrepancy $\left< \Delta e /e \right>$ [panel (b)], and density discrepancy $\left< |\Delta n| /\left|n \right| \right>$ [panel (c)].
 The results of the standard CNN with ReLU activation function are compared to the ones of the $\accnn$ ($N_c=260$). The vertical bars represents the standard deviation over the test set.
  }
\label{fig2}
\end{figure}

\begin{figure}[h!]
  \centering
  \subfigure{\includegraphics[scale=0.25]{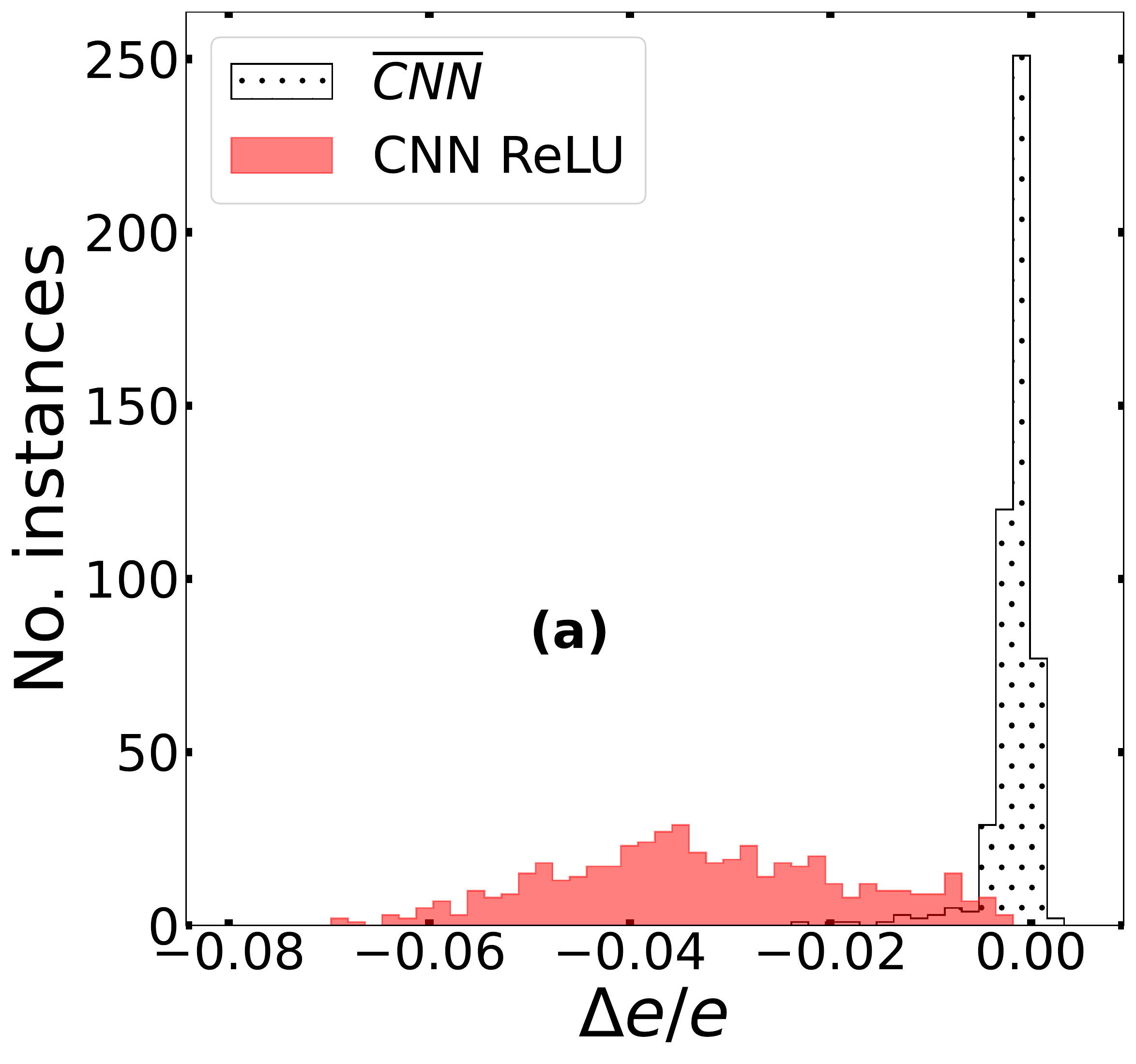} \label{e_hist_relu}}
  \hfill
  \subfigure{\includegraphics[scale=0.25]{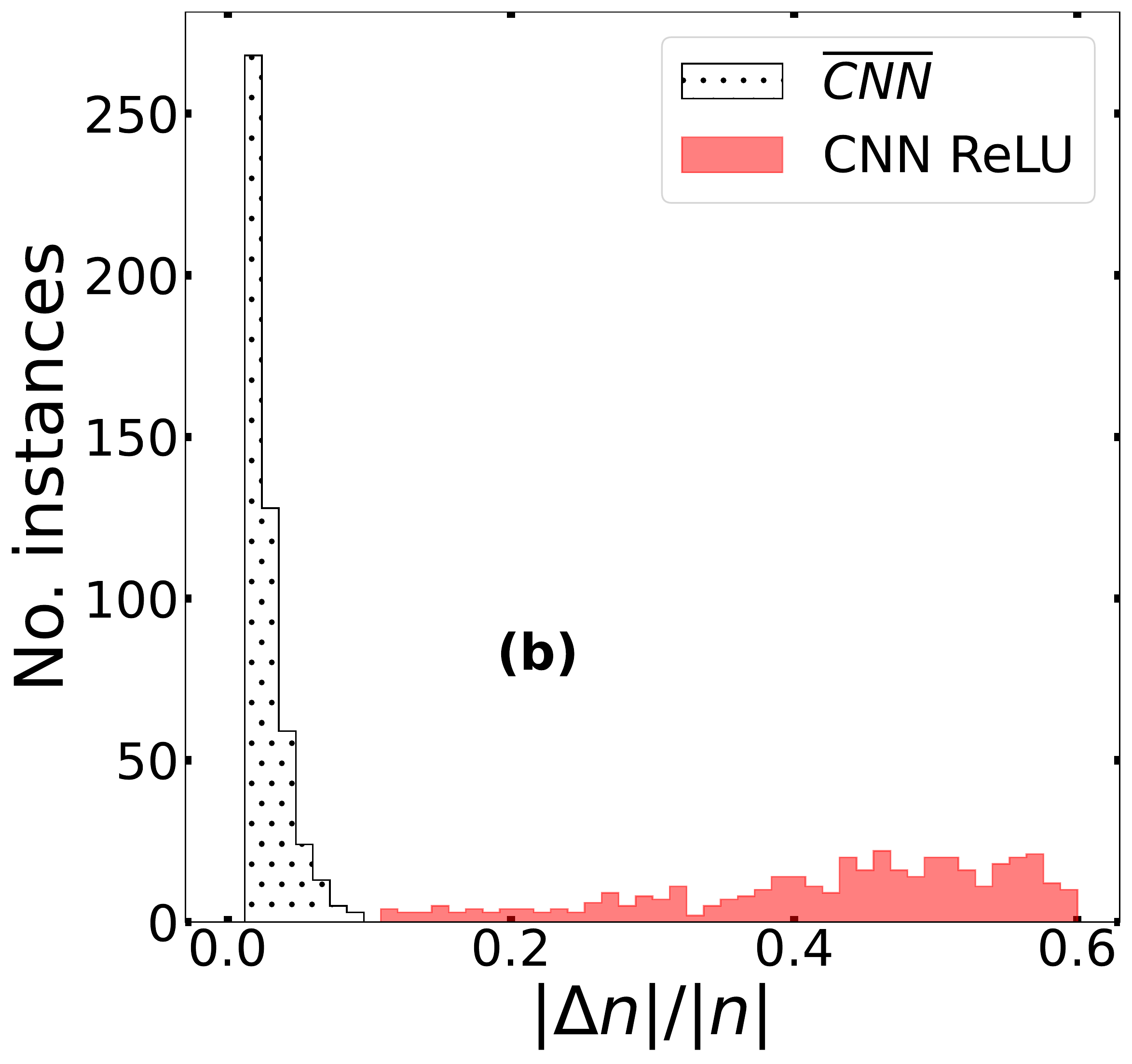}\label{n_hist_relu}}
 \caption{
Histograms of the relative energy discrepancies $\Delta e /e$ [panel (a)] and the density discrepancies $\left|\Delta n \right| /\left|n\right|$ [panel (b)] for a test set of 500 speckle-pattern instances, after $t_{\mathrm{max}}=10000$ steps of the gradient-descent optimization. 
The results of the standard CNN with ReLU activation function are compared to the ones of the $\accnn$.
 }
 \label{fig3}
\end{figure}

To be suitable for the DFT framework, the deep-learned functional $\tilde{T}{_\omega}[n]$ should allow iterating the gradient descent process as long as required to reach the minimum of $E[n]\equiv \tilde{T}{_\omega}[n] + \int \dd x V(x) n(x)$. 
Hereafter, we denote with $n_{\mm}(x)$ the density profile reached after gradient-descent optimization, and with $e_\mm=E[n_{\mm}(x)]$ the corresponding energy. The latter represents our estimate for the ground-state energy $e_\gs$.
Importantly, energies significantly lower than $e_\gs$ should never occur during the optimization process, as they would constitute a violation of the variational principle.
As explained in Section~\ref{sectionmethods}, the possible freezing in a local minimum significantly larger than $e_\gs$ can be circumvented by repeating the optimization from a different initial profile.
Figure~\ref{fig1} displays the density profiles reached after $t_{\mathrm{max}}=10000$  steps of gradient descent for two representative instances of the speckle pattern. 
For these and all other results reported below, the learning rate used in gradient descent is $\eta=10^{-3}$.
Clearly, the standard CNNs lead to unphysical profiles, while the $\accnn$ provides an accurate approximation of the exact profile $n_\gs(x)$.
To shed light on this phenomenon, we analyze the behavior of the energy discrepancy
\begin{equation}
    \Delta e= e_{\gs}- e_{\mm}
\end{equation}
and of the density discrepancy
\begin{equation}
    |\Delta n|= \sqrt{ \int \dd x  (n_{\gs}(x)-n_{\mm}(x))^2 },
\end{equation}
based on the $L_2$ metric
\begin{equation}
    \left| n \right|=\sqrt{\int \dd x  n^2(x)}
\end{equation}
along the gradient-descent process.
Specifically, we consider the average of the relative energy error $\left< \Delta e/e_{\gs} \right>$, 
of the relative absolute energy error $\left< |\Delta e|/e_{\gs} \right>$,
and of the relative density error $\left<| \Delta n|/|n_{\gs}| \right>$ computed over a test set of 500 instances.
Their dependence on the number of gradient-descent steps is shown in Fig.~\ref{fig2}. The vertical bars indicate the standard deviation over this test set, meaning that they represent the fluctuations among different realizations of the speckle pattern.
For both standard CNNs, after an initial decrease, the average absolute error increases. This means that the optimization process in not reliable, as it should be halted at an unknown intermediate number of steps. 
The average relative error becomes negative, indicating a violation of the variational principle.
The density error also increases after many steps, corresponding to the formation of unphysical density profiles with large spurious spatial fluctuations, as exemplified in Fig.~\ref{fig1}.
%
Instead, the absolute relative energy error corresponding to the $\accnn$ (with $N_c=260$ channels) systematically decreases until it saturates around a small value corresponding to $\sim 0.5\%$.
The average error saturates close to $\left< \Delta e/e_{\gs} \right> \sim 0$, meaning that significant violation of the variational principle do not occur.
The histograms shown in Fig.~\ref{fig3} compare the energy and the density errors obtained with the $\accnn$ and with the standard CNN with ReLU activation function (this outperforms the corresponding model with Softplus activation) after $t_{\mathrm{max}}=10000$ steps of the gradient descent optimization. The $\accnn$ energies are concentrated around zero error, while the standard CNN results are broadly distributed in the region of negative energy errors, corresponding to strong violations of the variational principle.

\begin{figure}[h!] 
  \centering
  \subfigure{\includegraphics[scale=0.25]{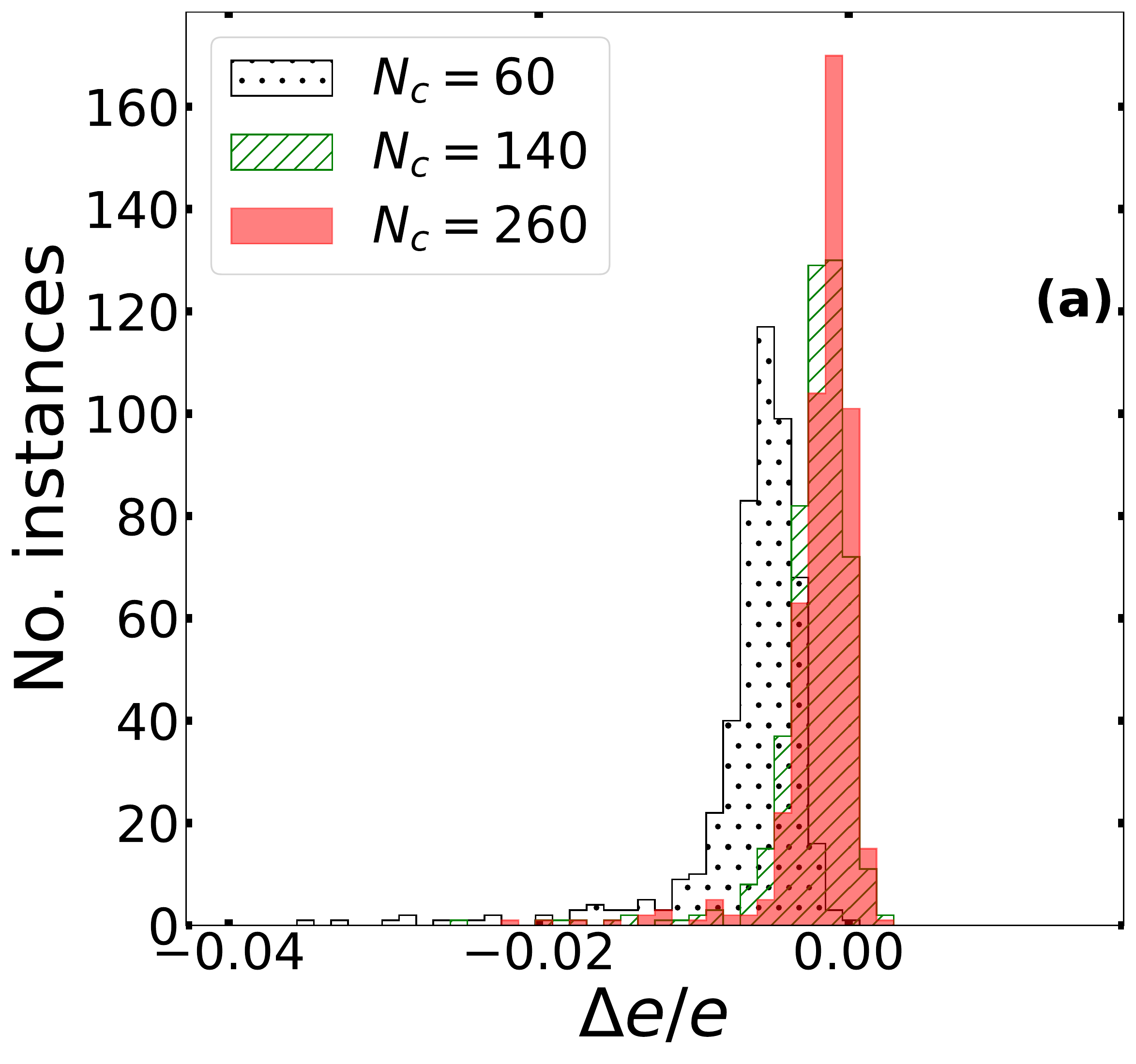}}
  \hfill
  \subfigure{\includegraphics[scale=0.25]{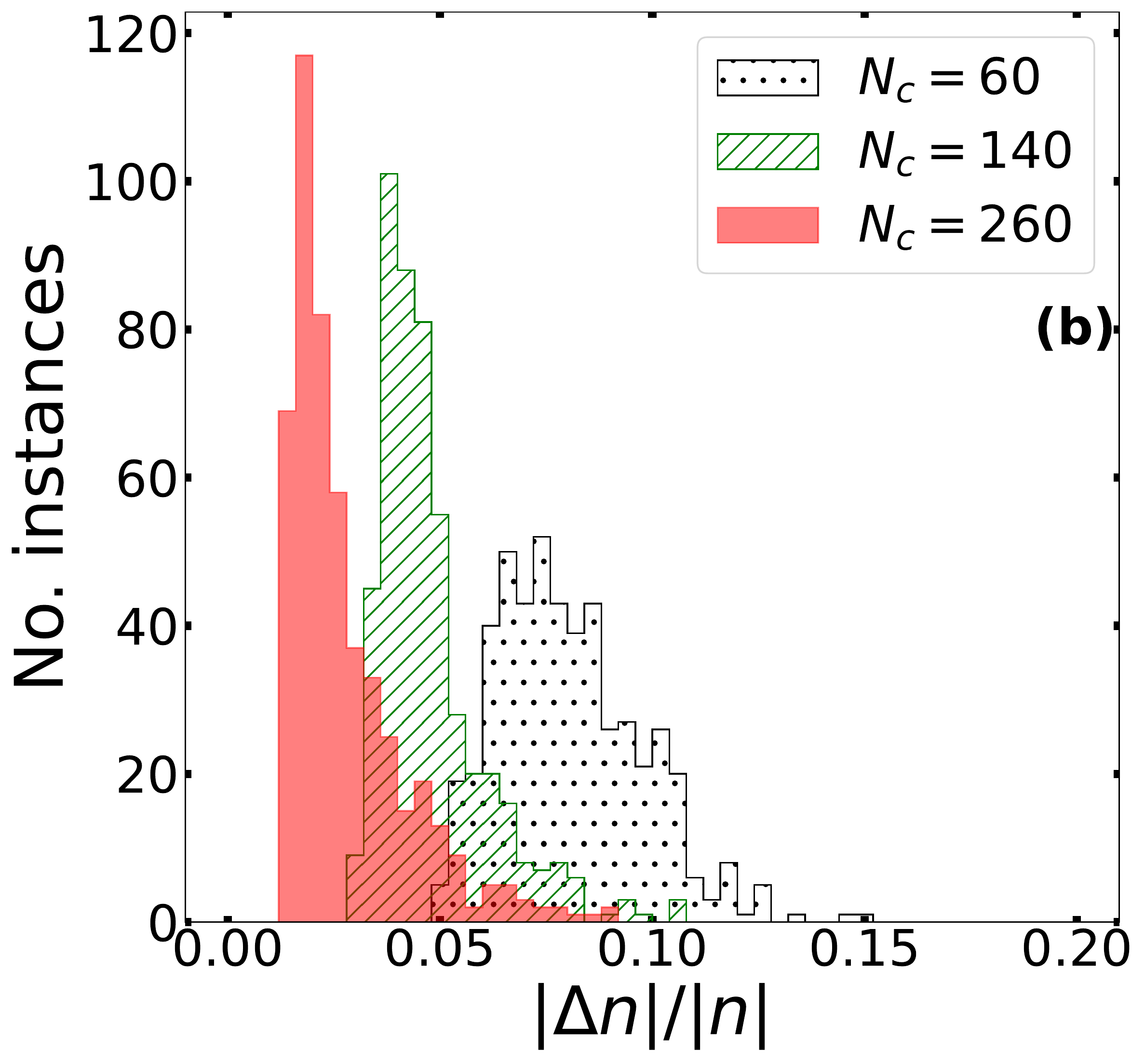} }
  \hfill
  \subfigure{\includegraphics[scale=0.25]{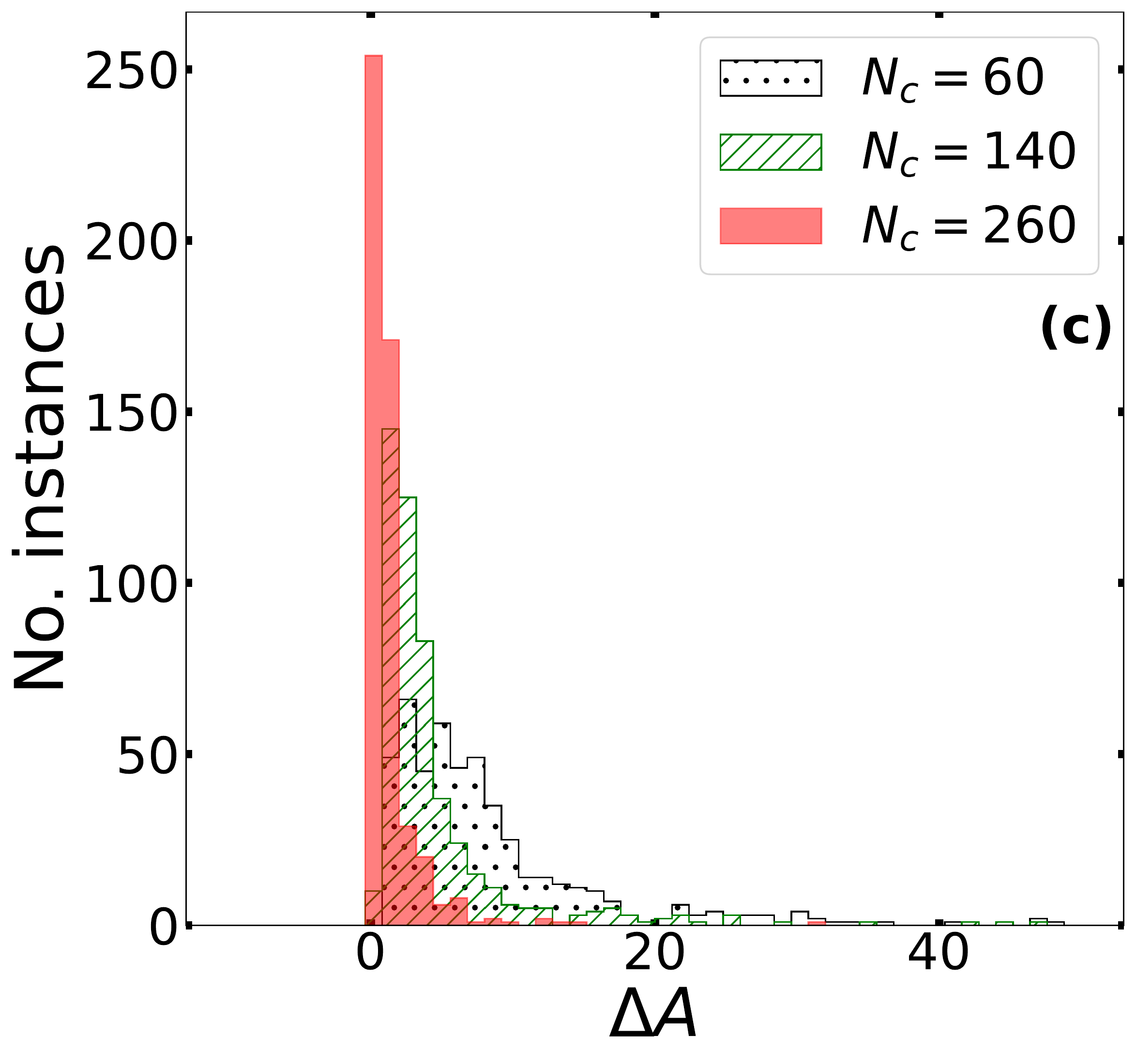} }
 \caption{
 Histograms of the relative energy discrepancies $\Delta e /e$ [panel (a)], of the density discrepancies $|\Delta n| /|n|$  [panel (b)], and of the noise metric $\Delta A$ defined in Eq.~\ref{eqnoise} [panel (c)], for a test set of 500 speckle-pattern instances, after $t_{\mathrm{max}}=10000$ steps of the gradient descent optimization. 
The results of the $\accnn s$ with different number of convolutional channels $N_c$ are shown.
}
 \label{fig4}
\end{figure}

Notably, the energy predictions obtained by performing gradient-descent optimization with the $\accnn$ systematically improve when the number of channels $N_c$ increases. This effect is shown in Fig.~\ref{fig4}. 
Notice that, for small $N_c$, small violations of the variational principle still occur in rare cases. However, they vanish for larger $N_c$. 
In fact, the average absolute energy discrepancy obtained after gradient descent with the $N_c=260$ $\accnn$ is $\left<|\Delta e|/e \right> \simeq 0.2 \%$, which is approximately twice the corresponding discrepancy obtained on a test set of exact grounds-state profiles $n_\gs(x)$, namely, $\left<|\Delta e_{ML}|/e \right> \simeq 0.1 \%$. 
Notice that this approximate doubling effect is expected, since the former error is also affected by the approximation in the density profile, while the latter corresponds to the  $\accnn$ prediction on the exact profile.
This means that gradient-descent optimization successfully identifies the ground state, within the residual uncertainty of the ML model.
Increasing $N_c$ leads also to more accurate density profiles (see panel (b) of Fig.~\ref{fig4}) and to the reduction of the spatial noise observed in the results provided by standard CNN (see Fig.~\ref{fig1} and also Refs.\cite{Meyer2020} \cite{PhysRevLett.108.253002}). %
%
%

%
%
%
%
%

%
%
%
To quantify this spatial noise, we consider the following metric:
\begin{equation}
\label{eqnoise}
    \Delta A= \frac{1}{N_{g}} \left< \sum_{i=1}^{N_{g}} \left( \left| \frac{  \nabla n_{\mm}(x_i)}{\nabla n_{\gs}(x_i)} \right| - 1 \right) \right>.
\end{equation}
It measures the error in the derivative of the density profile. Inaccurate profiles are characterized by large positive values $\Delta A \gg 1$, due to spurious spatial fluctuations, while exact predictions lead to $\Delta A =0$.
We find that large $N_c$ values lead to accurate and smooth density profiles (see panel (c) of Fig. \ref{fig4}), indicating the effectiveness of the average-channel layer. Residual local spurious fluctuations in the density profiles might be further suppressed via filtering procedures in post processing. Still, accurate DFT predictions are essential to avoid introducing biases by strong filtering procedures.

%
%

\begin{figure}[h!]
  \centering
  
  \subfigure{\includegraphics[scale=0.5]{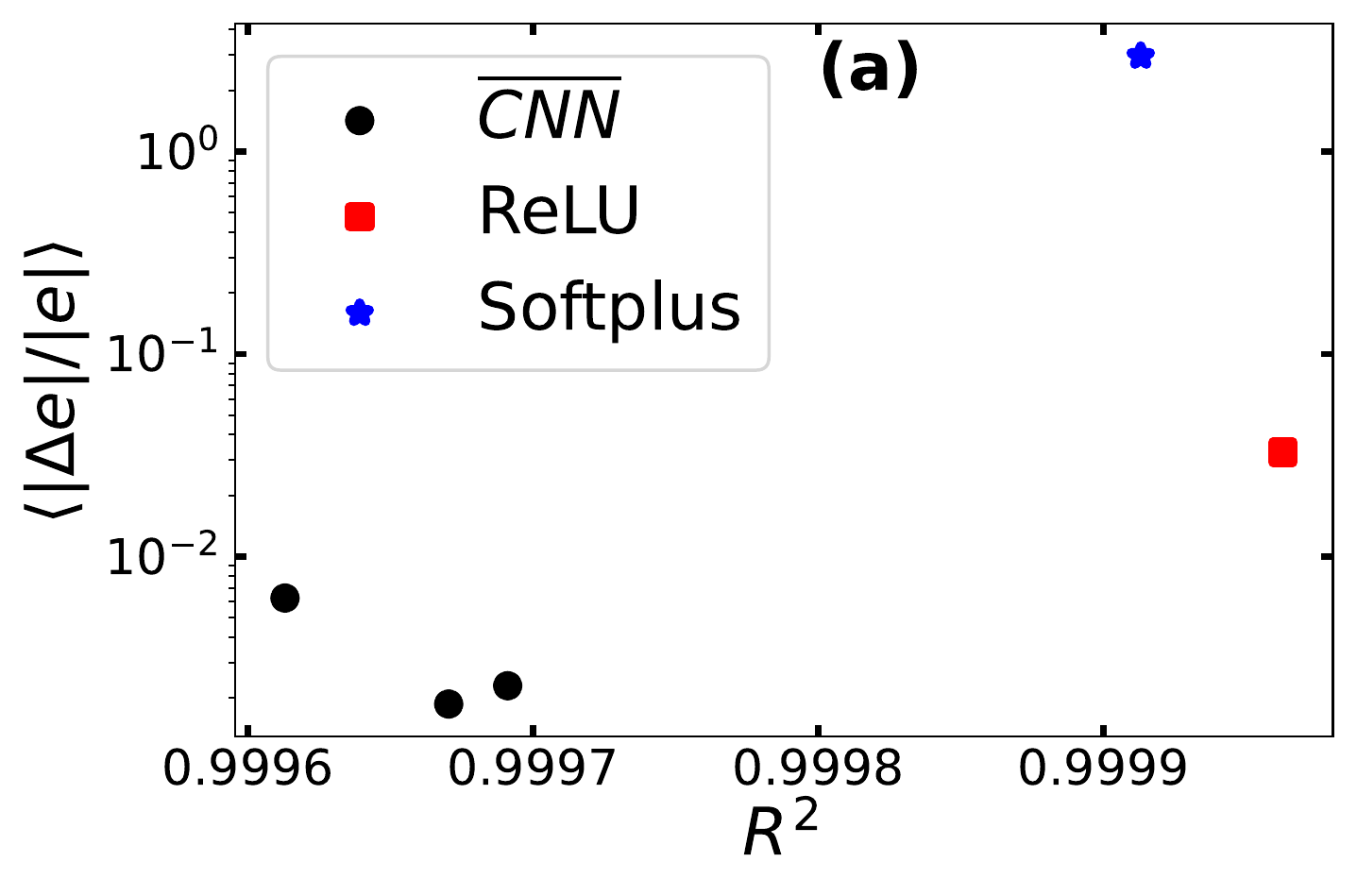}}
 \hfill
 \subfigure{\includegraphics[scale=0.5]{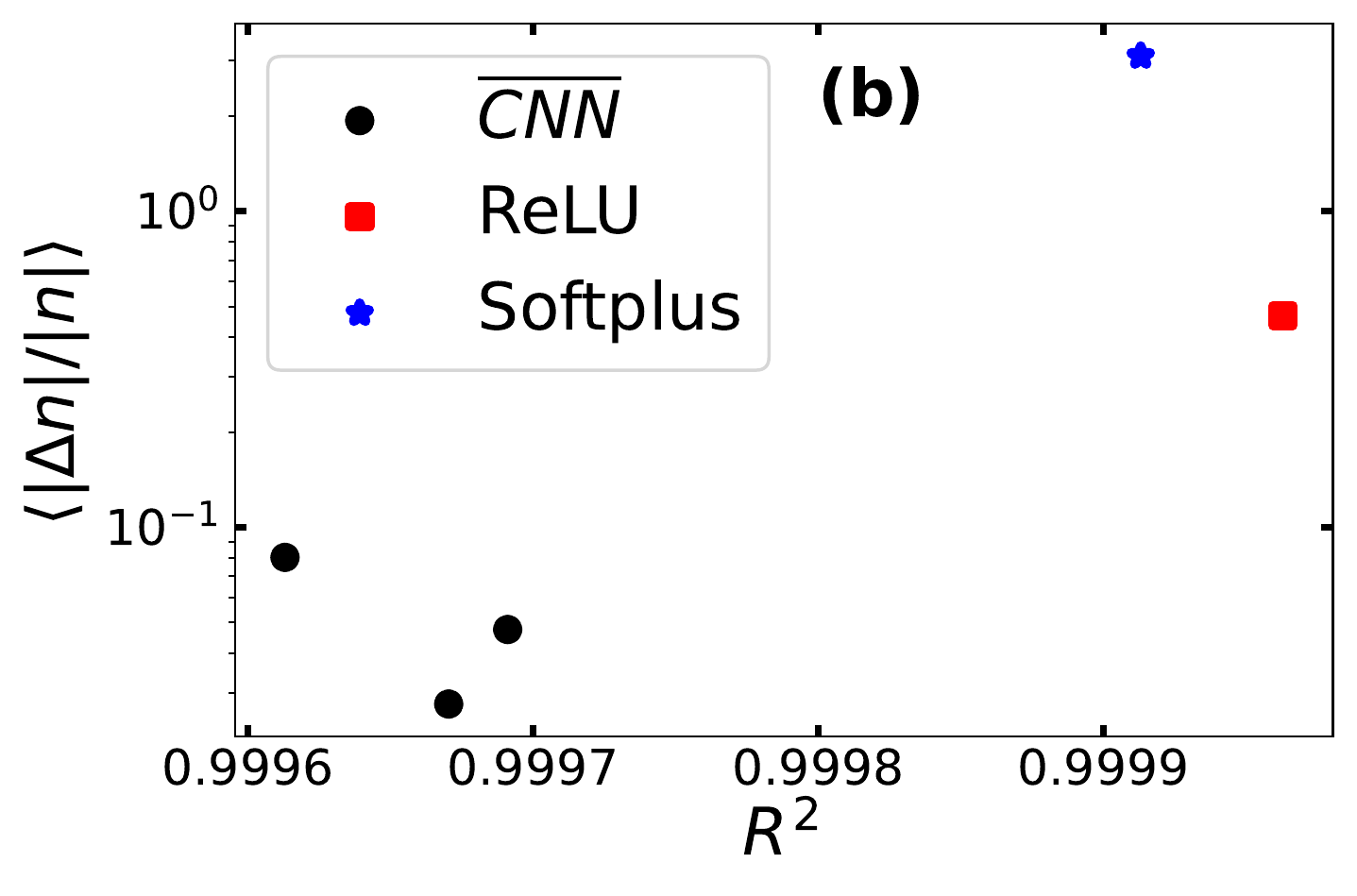} }
 \caption{
 Absolute relative discrepancy $\left< |\Delta e|/e \right> $ [panel (a)] and scatter plot of density discrepancy $\left<|\Delta n|/|n|\right>$  [panel (b)] after gradient-descent optimization ($t_{\mathrm{max}}=10000$, average over 500 instances), versus coefficient of determination $R^2$ over a test set of exact ground-state densities. Three neural networks are compared: standard CNN with ReLU and with Sofplus activation functions, and $\accnn$.
 }
 \label{fig5}
\end{figure}

It is worth further emphasizing that using the standard metrics of deep learning on training and test sets of exact density profiles is not necessarily helpful to predict the performance of the ML functional in the variational minimization of DFT. 
This is exemplified by the scatter plots of Fig.~\ref{fig5}. They display the errors obtained after gradient descent optimization versus the coefficient of determination Eq.\eqref{R2} computed on a test set of exact ground-state density profiles. The two standard CNNs and the $\accnn$ with three values of $N_c$ are considered. No (anti) correlation is clearly noticeable, meaning that high prediction accuracies on exact ground-states do not necessarily correspond to highly effective functionals for DFT. This should be taken into account in the future development of deep learning techniques for DFT.

\section{Conclusions}
\label{conclusions}
The progress in data-based DFT is currently being hindered by the instabilities encountered when using ML functionals in gradient-based optimization.
We presented a promising approach to circumvent this problem. This relies on the implementation of a deep neural network  tailored to DFT. Specifically, we have shown that when  an inter-channel averaging layer is included, beyond the standard convolutional, pooling, and dense layers, gradient-descent optimization can be iterated at will, obtaining accurate ground-state energies and density profiles and avoiding violations of the variational principle beyond residual uncertainties from the imperfect training of the regression model.
Our analysis has focused on a realistic one-dimensional model for non-interacting atoms in optical speckle disorder, which leads to rather variegate density profiles compared to models addressed in previous studies on ML based DFT.
For completeness, in the Appendix~\ref{appendix} the performance of the $\accnn$ in Gaussian-well models borrowed from the literature is demonstrated. On the one hand, this further analysis indicates the rather general range of applicability of our tailored neural network. On the other hand, it points out the need of training sets including significantly variegate density profiles.
To favor future comparative studies, our training datasets are made freely available at Ref.~\cite{costa_emanuele_2022_6504567}. 

Additional challenges are going to be faced in the further development of ML techniques for DFT~\cite{commentBurke}.
A further assessment of the network effectiveness should focus on higher dimensional and interacting models. In the DFT formalism, moving from single-particle to many-body problems is less challenging than in wave-function based methods, since observables are still obtained from the single-particle density. Therefore, we expect the $\accnn$ to be useful also in the many-body context. Clearly, generating training datasets is, in that case, more 
demanding~\cite{2022,tamblynDMC}. The learning process has to be accelerated, and the following strategies could be adopted.
Incorporating physics knowledge into the deep-learning framework is a possible strategy~\cite{nagai2020completing,PhysRevLett.126.036401}.
Another promising approach is transfer learning. This technique has already proven suitable to accelerate the supervised learning of the ground-state properties of both non-interacting and interacting quantum systems~\cite{mills2019extensive,saraceni2020scalable,10.21468/SciPostPhys.10.3.073}. Interestingly, even extrapolations were proven feasible, meaning that (scalable) networks trained on relatively small systems provided accurate predictions for larger sizes or larger particle numbers. These techniques might be adopted also in the framework of DFT. We leave these endeavors to future investigations.

\section*{Acknowledgments} This work was supported by the Italian Ministry of University and Research under the PRIN2017 project CEnTraL 20172H2SC4. 
S.P. acknowledges PRACE for awarding access to the Fenix Infrastructure resources at Cineca, which are partially funded by the European Union’s Horizon 2020 research and innovation program through the ICEI project under the Grant Agreement No. 800858. 
S. P. also acknowledges the Cineca award under the ISCRA initiative, for the availability of high performance computing resources and support. 

\appendix
\section{Gaussian-well potentials}
\label{appendix}
\begin{figure}[h!]
  \centering
  \subfigure{\includegraphics[scale=0.25]{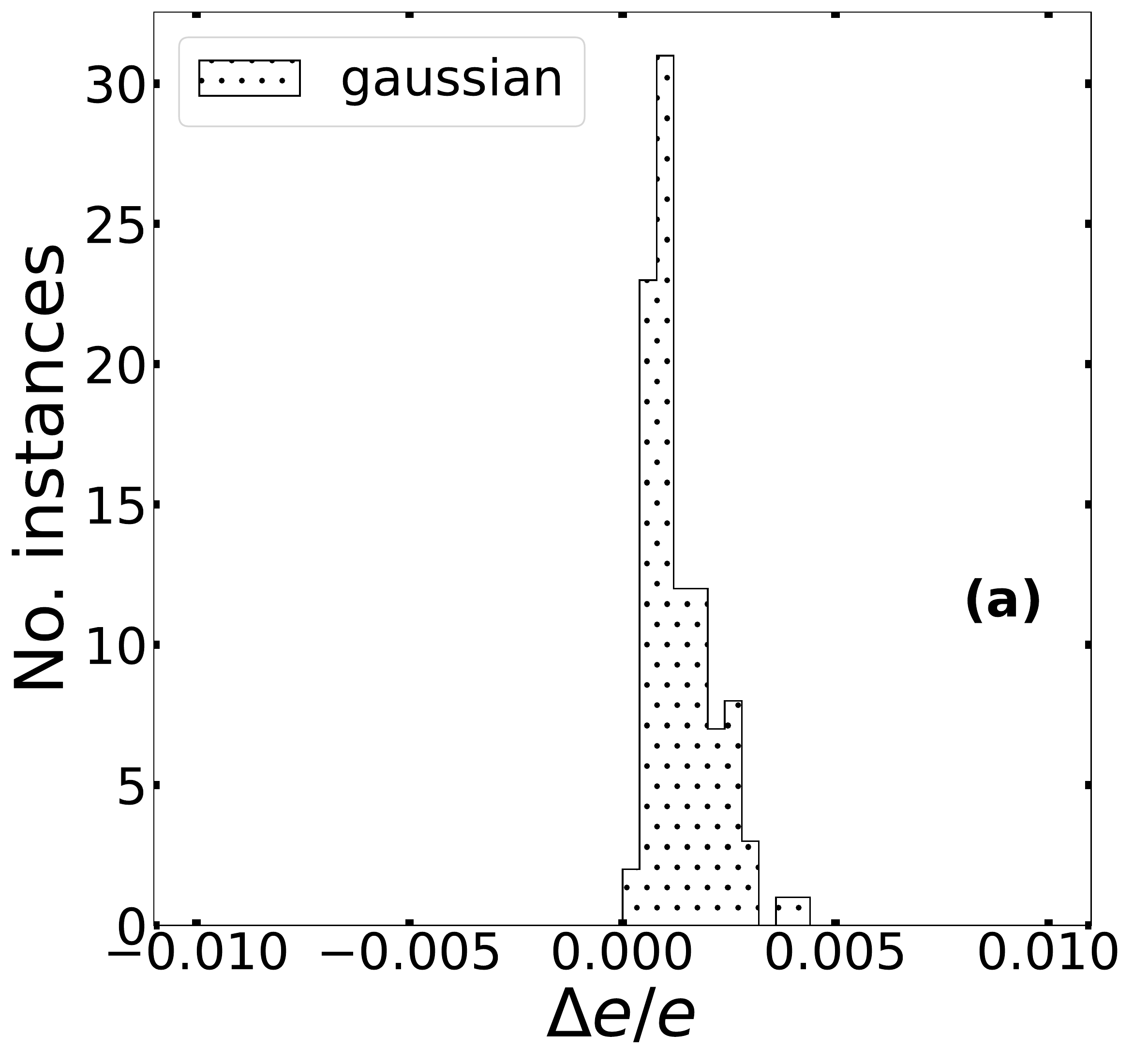} \label{e_hist_gau}}
  \hfill
  \subfigure{\includegraphics[scale=0.25]{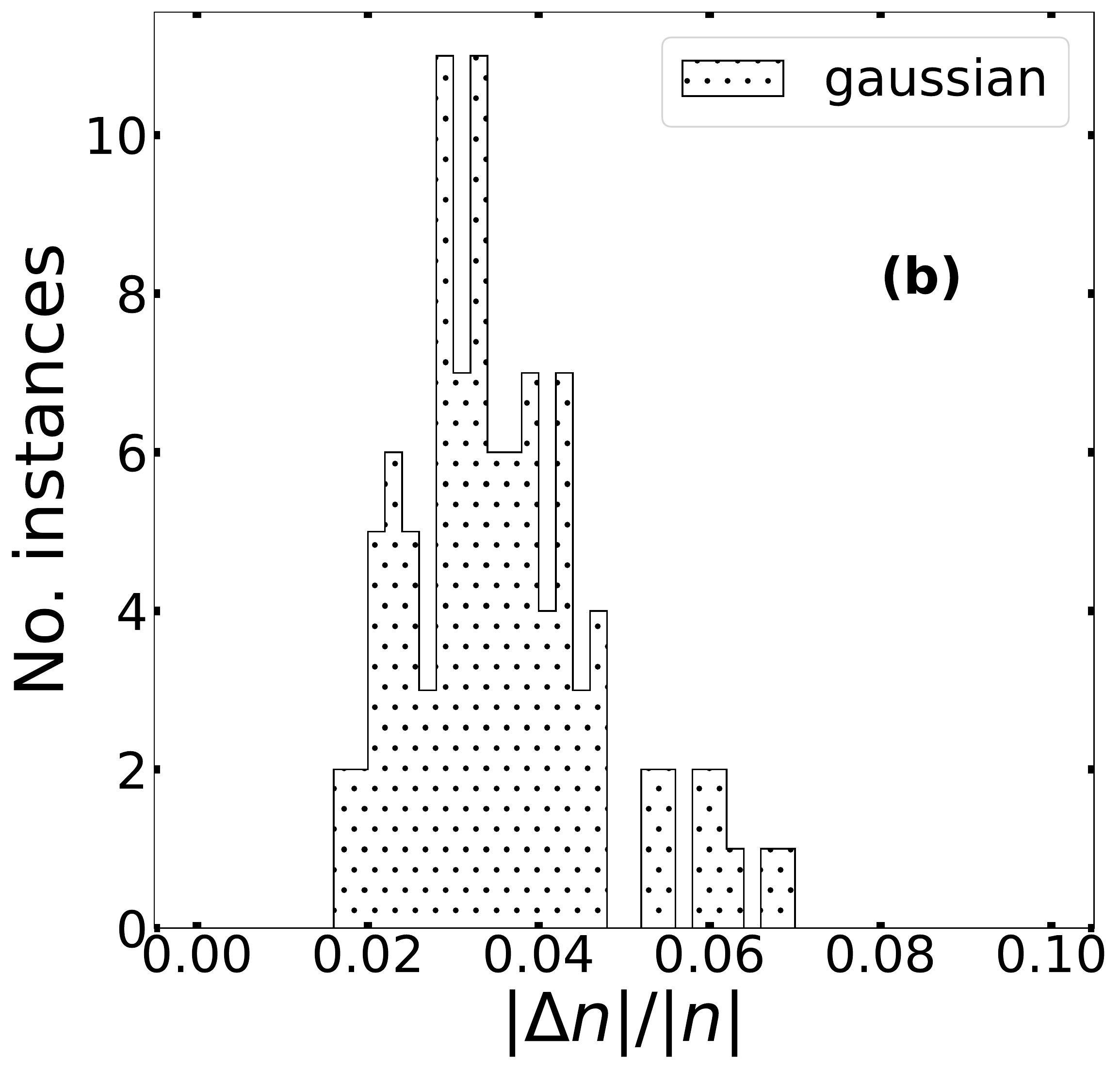}\label{n_hist_gau}}
 \caption{
Histograms of the relative energy discrepancies $\Delta e /e$ [panel (a)] and the density discrepancies $\left|\Delta n \right| /\left|n\right|$ [panel (b)] for a test set of $100$ Gaussian-well potentials, after $t_{\mathrm{max}}=10000$ steps of the gradient-descent optimization. 
 }
 \label{fig6}
\end{figure}

\begin{figure}[h!]
  \centering
  \subfigure{\includegraphics[scale=0.15]{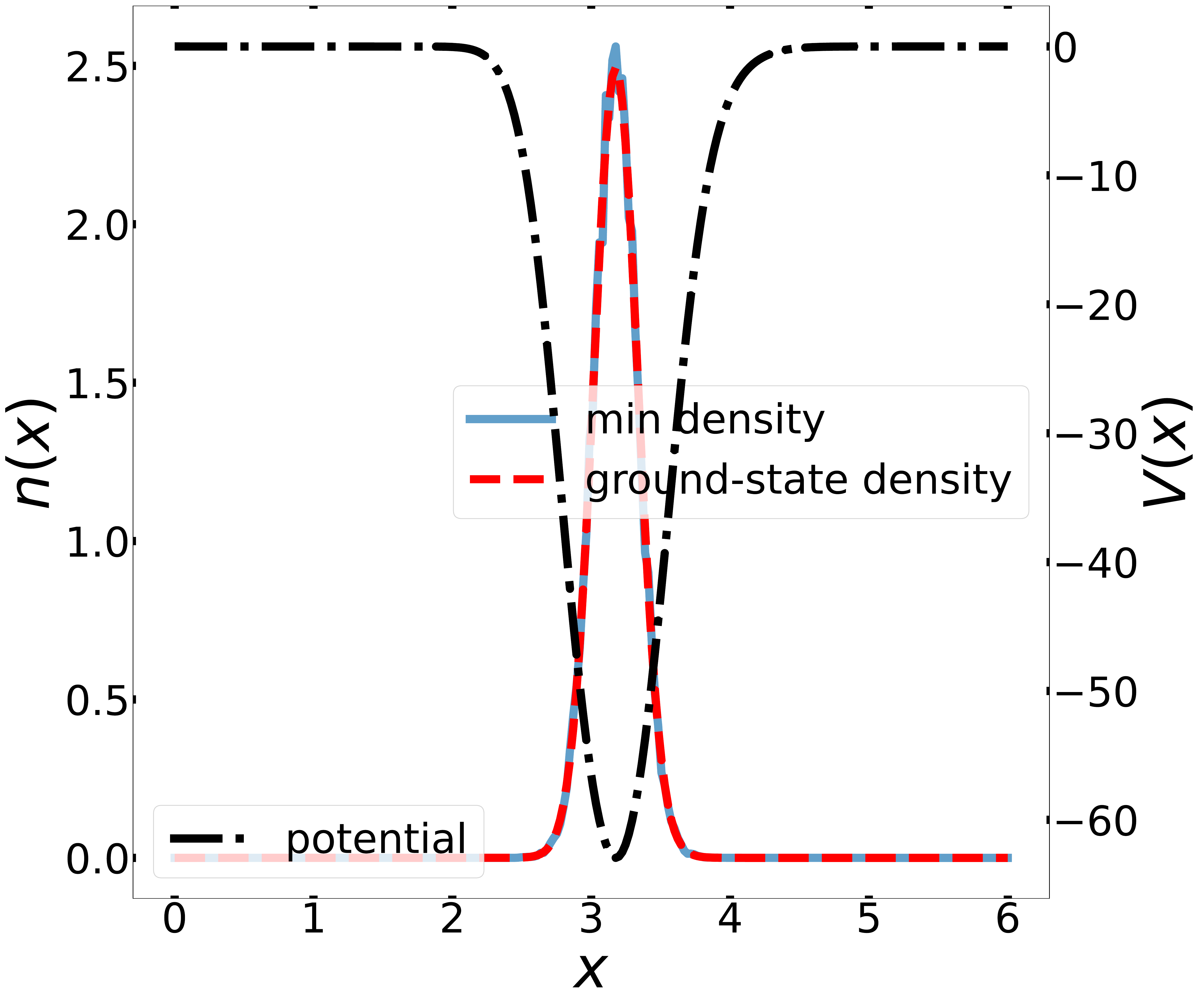}}
  \caption{
 External potential $V(x)$ corresponding  to an instance of the Gaussian-well potential (right vertical axis, unit of  $\frac{h^2}{m \ell^2}$) and the corresponding ground-state density profiles (left vertical axis, units of $1/\ell$). $\ell$ is the length unit, used on the horizontal axis. The density configuration  obtained by gradient-descent minimization is compared to the exact density profile.
  }
\label{fig7}
\end{figure}

To favor direct comparison, and to further characterize the effectiveness of the $\accnn$, here we consider a different testbed model, borrowed from previous studies. It describes narrow potential wells which confine the particle in the central region of a finite box. Setting the units such that  $\hbar^2/(m\ell^2)=1$, where $\ell$ is the length unit, the (adimensional) single particle Hamiltonian reads:
\begin{equation}
    H= -\frac{1}{2} \frac{\mathrm{d}^2}{\mathrm{d} x^2} + V(x),
\end{equation}
with the external potential
\begin{equation}
    V(x)=-\sum^{3}_{i=1} a_i \exp\left[-\frac{(x-b_i)^2}{2c^{2}_i}\right].
\end{equation}
The model parameters $a_i$, $b_i$, and $c_i$ are sampled from uniform probability distributions in the following ranges: $a_i \in [1,40]$, $b_i \in [1.8, 4.2]$, and $c_i \in [0.12,0.4]$. The box size is $L=6$, and hard-wall boundary conditions are adopted. The chosen box is $L=6$, rather than $L=1$ as in Ref.\cite{Meyer2020},  so that the ground-state density profiles essentially vanish before reaching the boundaries. This strongly suppresses the role of the choice of the boundary conditions. Indeed, we find negligible variations in the ground-state energies with hard-wall compared to periodic boundary conditions. With the size $L=1$, the boundary effects are sizable, and the density profiles corresponding to different potential instances display small variations when hard-wall boundaries are adopted. This does not allow an effective training of the deep neural network, reintroducing instabilities in the gradient-descent optimization. We find that this problem is solved either enlarging the system size, e.g., to $L=6$, as shown hereafter, or by adopting periodic boundary conditions in a small box, as mentioned later on in this paragraph.

A $\accnn$ is trained on a global dataset of $150000$ instances ($90 \%$ train, $10 \%$ validation), using the same structure ($2$ blocks, pooling size $[4,2]$, kernel size $13$, and $260$ hidden channels) and the same hyperparameters as in the case of the speckle potential with a number of epochs $N_e=3000$. However, here the convolutional layers use zero padding, as opposed to the periodic padding adopted for the speckle potentials, which are defined within periodic boundary conditions.
The gradient descent performance on a test set of 100 instances is visualized in Fig.~\ref{fig6}. Again, we find that gradient-descent optimization ($t_{\mathrm{max}}=10000$) leads to accurate ground-state energies ($\left<|\Delta e|/e \right> \simeq 0.14 \%$) and density profiles ($\left<|\Delta n|/|n| \right> \simeq 3 \%$ ), without sizable violation of the variational condition.
%
%
%
%
For illustrative purposes, an example of Gaussian-well potential, with the corresponding density profile obtained with $L=6$, is shown in Fig.~\ref{fig7}.
To facilitate comparison, we report the model details and the performance metrics using the units and the conventions adopted in some previous studies. These considered one-electron models in atomic units, so that $\ell$ corresponds to the Bohr radius and the energy unit $\hbar^2/(m\ell^2)$ corresponds to one Hartree (Ha). The mean  kinetic energy of the $L=6$ model is $\bar{t}_{\mathrm{gs}} \simeq 3.994 \mathrm{Ha}$ and the standard deviation is quite large, namely $\sigma\simeq 1.457 \mathrm{Ha}$, with maximum value $ 12.957 \mathrm{Ha}$ and minimum value $ 0.2605 \mathrm{Ha}$, corresponding to considerably variegate ground states. 
The average kinetic-energy prediction-error on a test set of $N_{\mathrm{test}}=15000$ instances is $\left<\left| \Delta t_{\mathrm{ML}}\right|\right> \equiv N_{\mathrm{test}}^{-1}\sum_{k=1}^{N_{\mathrm{test}}}
    \left|t_{\gs,k}-\tilde{T}_{\omega}[n_{\gs,k}]\right|
     \simeq 8 \cdot 10^{-4} \mathrm{Ha}$; the standard deviation of $\left| \Delta t_{\mathrm{ML}} \right|$ is $\sigma \simeq 11 \cdot 10^{-4} \mathrm{Ha}$.
After gradient descent, the average absolute kinetic-energy error is $\left< \left| \tilde{T}_{\omega}[n_{\mathrm{min}}]-t_{\mathrm{gs}}\right|\right> \simeq 0.0171 \mathrm{Ha}$ and the corresponding standard deviation $\sigma \simeq 0.0191 \mathrm{Ha}$.
The average absolute density discrepancy is $\left< \left| \Delta n \right| \right>=0.031$ and the corresponding standard deviation is $\sigma \simeq 0.008$.

It is also worth mentioning that, considering $L=1$ and periodic boundary conditions, the accuracy metrics
are comparable as above, namely: $\left<|\Delta e|/e \right> \simeq 0.3 \%$ and $\left<|\Delta n|/|n| \right> \simeq 3 \%$. In this case, periodic padding is used, and the first left and right periodic images of the Gaussian wells are included to make the potential $V(r)$ essentially periodic. 
The potential parameters are sampled in the following ranges $b_i\in [0.1,0.9]$, $a_i \in [1,40]$ and $c_i\in [0.03,0.1]$. When combined with periodic boundary conditions, these allow creating sufficiently variegate density profiles despite of the small system size.
These findings indicate that the $\accnn$ represents a flexible regression model to apply ML-based DFT to rather arbitrary external potentials, whereby the density profiles display significant variations among different random realization of the sample.

\bibliography{main}{}

\end{document}